# High resolution miniature dilatometer based on AFM piezocantilever

J.-H. Park, D. Graf, T. P. Murphy, and S. W. Tozer

*National High Magnetic Field Laboratory, Florida State University, Tallahassee, Florida 32310*

G. M. Schmiedeshoff

*Department of Physics, Occidental College, Los Angeles, California 90041*

**Abstract**

Thermal expansion, or dilation, is closely related to the specific heat, and provides useful information regarding material properties. The accurate measurement of dilation in confined spaces coupled with other limiting experimental environments such as low temperatures and rapidly changing high magnetic fields requires a new sensitive millimeter size dilatometer that has little or no temperature and field dependence. We have designed an ultra compact dilatometer using an atomic force microscope (AFM) piezoresistive cantilever as the sensing element and demonstrated its versatility by studying the charge density waves (CDWs) in alpha uranium to high magnetic fields (up to 31 T). The performance of this piezoresistive dilatometer was comparable to that of a titanium capacitive dilatometer.

When probing the physical properties of condensed matter, the practical size of a sample and the dimensions of the instrument can limit the experimental accuracy and the parameter space. Typically crystals of exotic new materials are so small that even attaching four wires can be a daunting task. Measuring the dilation of small samples in high magnetic fields and at low temperatures, i.e. magnetostriction, is even more difficult as the signal is directly proportional to the length of the sample. Previously we introduced a versatile compact metallic capacitive dilatometer with a resolution better than 0.1 Å [1]. The device measures a change in capacitance between two parallel plates that is a function of sample dilation due to changes in magnetic field or temperature. However, when the sample is submerged into a liquid or gas cryogenic medium for heat exchange, the device measures the signal from the sample dimension change and the change of dielectric property of the cryogenic medium itself, complicating data analysis. There are other experimental issues with capacitance cells such as the influence of gravity on the floating plate when the cell is rotated, eddy current heating at millikelvin temperatures induced during field sweeps, cell effects that arise from the change in the instrument's own dimension rather than that of the sample, uniaxial stress on the sample and a temperature lag between the sample and the temperature sensor. Neumeier et al. [2] have addressed the issue of eddy current heating and the cell effect by making a capacitance dilatometer from fused quartz, but the size of the cell makes it difficult to rotate at cryogenic temperatures. We have also fabricated a smaller version of our metallic cell that rotates in a 14 mm diameter space and minimizes eddy current heating since it is made from more resistive metals than the copper original. In this report, we present an ultra compact piezoresistive dilatometer that overcomes all of the aforementioned issues.

The principle of operation is to measure the change in resistance of an AFM piezoresistive cantilever (PRC) when the sample dimensions change. As schematically shown in Fig. 1, for a z-direction dilation measurement, the sample and PRC are glued to a substrate. The tip of the PRC gently rests on the sample with the AFM tip facing up such that a 1% change in nominal resistance is generated (this value was determined and adjusted via trial and error) thereby assuring that the tip and sample will not separate as the sample contracts upon cooling.

The resistances of the piezo element and the reference piezo element then can be monitored using a Wheatstone bridge configuration. Resistance changes in the piezo element induced by dimensional changes in the sample are recorded as a voltage change in a lock-in amplifier. With this simple idea, we have built the first prototype PRC dilatometer (Fig. 2) using a commercial atomic force microscope (AFM) piezoresistive cantilever [3, 4]. Because of its extreme sensitivity to mechanical movement and low cost this piezoresistive cantilever has also found use in recent ultra sensitive magnetic torque measurements [5].

As a test sample, a single crystal of depleted alpha uranium (sample #1, thickness 0.04 mm) was glued to the upper part of a stepped silicon substrate using commercial super glue. By matching the cantilever and substrate material, we have been able to minimize the cell effect. The height of the step is designed such that the tip of the cantilever is gently resting on top of the sample with the AFM tip facing up as described previously. The dimension of the commercial PRC device measures 3.5 mm x 1.6 mm x 0.25 mm (L x W x H) overall and lever arm has dimensions of 0.4 mm x 0.05 mm x 0.005 mm (L x W x H). A typical room temperature resistance of the piezo element ($R_{ab}$ or $R_{cd}$) was about 625 ohms. A low frequency (~ 10 Hz) bias voltage (0.1 V) was used as a source for DC magnetic field measurements (see also Ref. 5 for pulsed field application configuration). When necessary, a low frequency filter with preamplifier was used for the signal detection.

For a comparison study, another thicker crystal (sample #2, thickness 0.4 mm) was mounted in the 19 mm diameter capacitive titanium dilatometer and the PRC dilatometer was placed on top of the titanium dilatometer housing, see Fig. 2a. The two dilatometers were then installed on the probe of the continuously flowing helium variable temperature insert and the dilations were measured upon warming from 15 K to 50 K at 2 K/min for fields up to 31 T.

One of the initial purposes of the experiment, aside from testing a new dilatometer, was to see if the charge density waves (CDWs) in alpha uranium could be detected and then suppressed with magnetic field. Alpha uranium [6] is one of the few elements that goes through a sequence of CDW transitions (denoted as $\alpha_1$, $\alpha_2$, and $\alpha_3$ in the resistivity plot, Fig. 3) and since the CDW is closely related to the lattice structure, the dilatometer should be an ideal instrument to study CDW transitions.

Figure 3 (top panel) shows a resistance trace of typical alpha uranium with CDW transitions at $\alpha_1$ (43 K), $\alpha_2$ (37 K), and $\alpha_3$ (23 K). Similar CDW transitions were observed but in a more pronounced manner in the PRC dilatometer study (bottom panel). A detailed analysis of the sensitivity of the PRC dilatometer requires a suitable calibration procedure (to calculate change in sample dimension from the resistance change of PRC piezo-elements) which is still under investigation. However, initial measurements on samples of the same material, from the same source, over the same range of temperatures (15 K ~ 50 K) and magnetic fields (0 ~ 31 T), suggest that the sensitivity of our cantilever design is comparable to that of our capacitive design [1], and thus, the use of this new PRC dilatometer holds great promise in the area of DC and pulsed high magnetic field measurements that are often space limited.

In fact, in our preliminary experiments using the PRC dilatometer in pulsed magnetic field, we have observed a metal to insulator (MI) transition in a organic compound, $\tau$-(P-(r)-DMEDT-TTF)$_2$(AuBr$_2$)$_{1+y}$ and the hysteric behavior of the MI transitions identified from the previous study [7] was reproducible (see Fig 4a) at a field of approximately 36 T at a temperature of 1.7 K. We also have observed quantum oscillations in the single crystal LaRhIn$_5$ in DC magnetic field, as shown in Fig 4b and find all the frequencies measured by Shishido et al. [8] and the theoretically predicted $\varepsilon_1$ frequency, see Fig 4c. The frequencies of quantum oscillations were reproducible in the up and down field sweeps.

For immediate future improvement the following are under consideration. To calibrate the PRC dilatometer for absolute measurements, a known standard sample can be measured. Alternatively, an encoded linear piezo drive can be used to generate calibrated dimension changes. Since the dimensions of the dilatometer are small, it can easily be rotated or used in the confined space of a pressure cell. Simultaneous measurements of sample dilation and heat capacity may also be possible with a cantilever mounted on a commercial heat capacity puck, the small thermal mass of the cantilever contributing to the heat capacity addenda. The small size may also permit measurement of more than one sample dimension simultaneously using multiple cantilevers.

The authors would like to thank Jim Willit of Argonne National Laboratory for the alpha uranium samples, John Sarrao and Pascoal Pagliuso of Los Alamos National Laboratory for the LaRhIn$_5$ sample, and G.C. Papavassiliou of the National Hellenic Research Foundation for the $\tau$-(P-(r)-DMEDT-TTF)$_2$(AuBr2)$_{1+y}$ sample. Support for this work was provided by the DOE/NNSA under DE-FG52-06NA26193. Work was performed at the National High Magnetic Field Laboratory which is supported by NSF Cooperative Agreement No. DMR-0654118 and by the State of Florida. Work at Occidental College was supported by the NSF under DMR-0704406

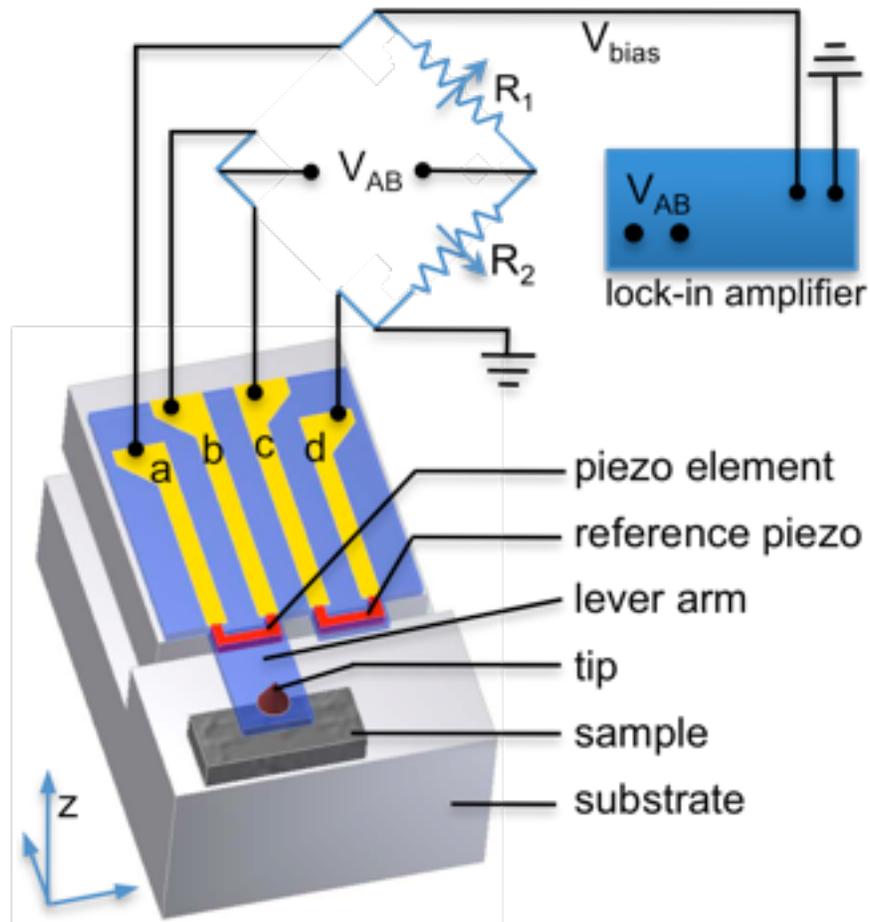

Fig. 1. Schematic diagram of piezoresistive cantilever dilatometer. A commercial piezoresistive cantilever for AFM is mounted on a stepped silicon substrate with its tip resting on the sample that is glued on the upper part of the substrate. The resistance change between the sample piezo and reference piezo elements due to the dimensional change in sample is monitored using a Wheatstone bridge technique.

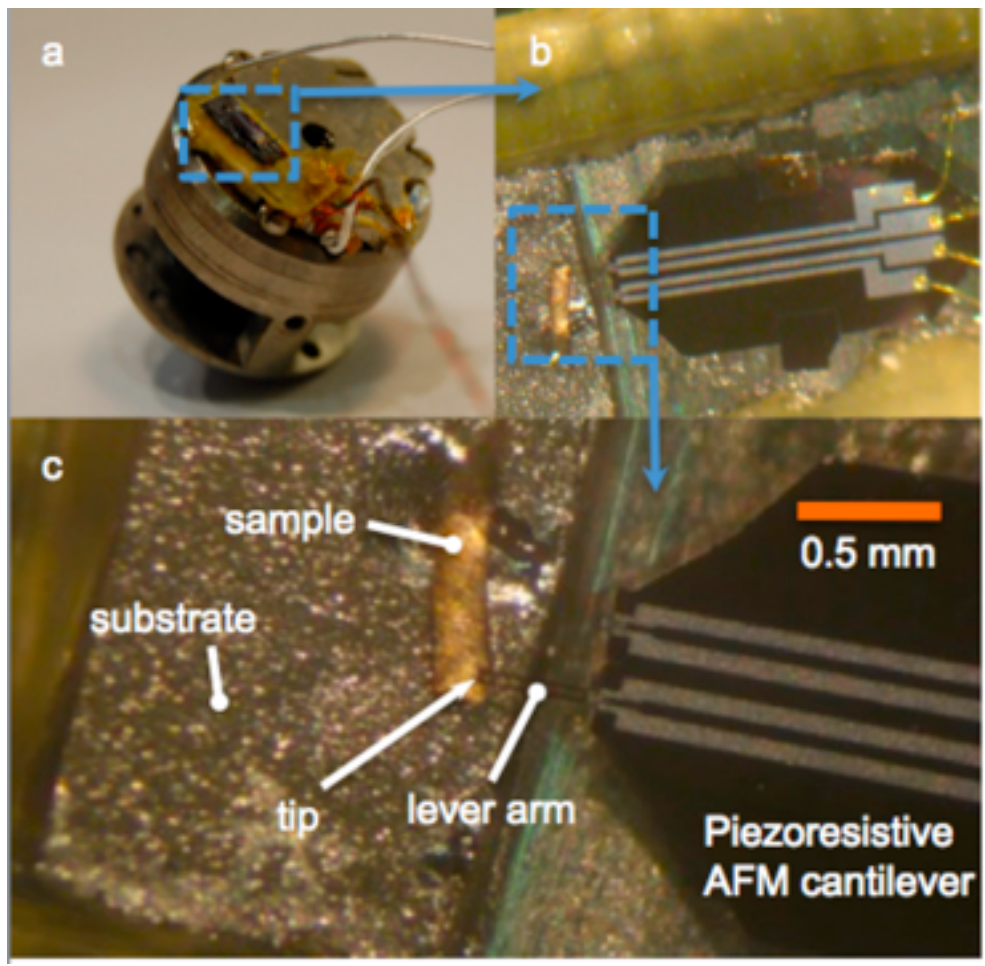

Fig. 2. Piezoresistive dilatometer placed on top of the capacitive titanium dilatometer with a diameter of 19.0 mm (a). 2b and c show details of the piezoresistive dilatometer and sample mounting.

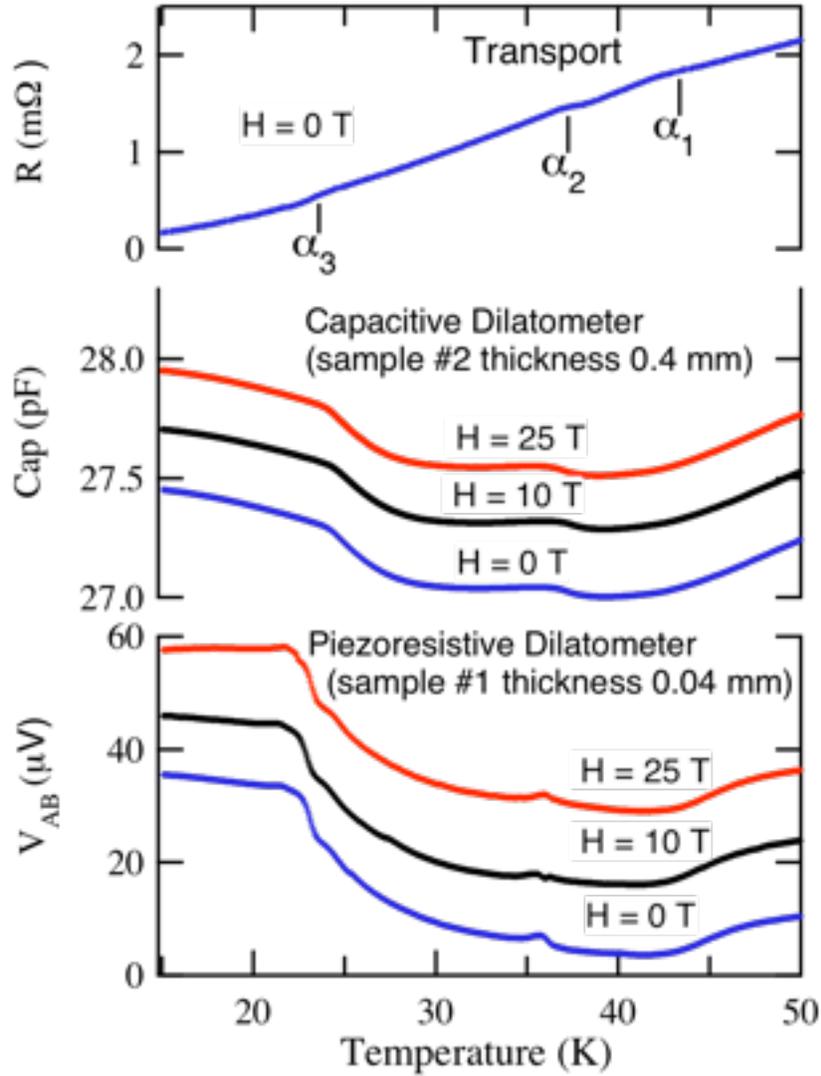

Fig. 3. Charge density wave transitions in alpha-uranium probed by transport (top panel), capacitive dilatometer (middle), and piezoresistive dilatometer (bottom). The dilatometry measurements were taken simultaneously with the PRC dilatometer glued to the top of the capacitance dilatometer as shown in Fig. 2. Dilatometer traces at different magnetic fields are not corrected for cell effects and offset along the y-axis for clarity. The $V_{AB}$ of piezoresistive dilatometer (bottom) was measured without a preamplifier. The dilations (in both dilatometers) measured at 31 T (not shown here) also showed similar CDW transition temperatures. *Note that the thickness of sample in capacitive dilatometer is 10 times thicker than that of the piezoresistive dilatometer*.

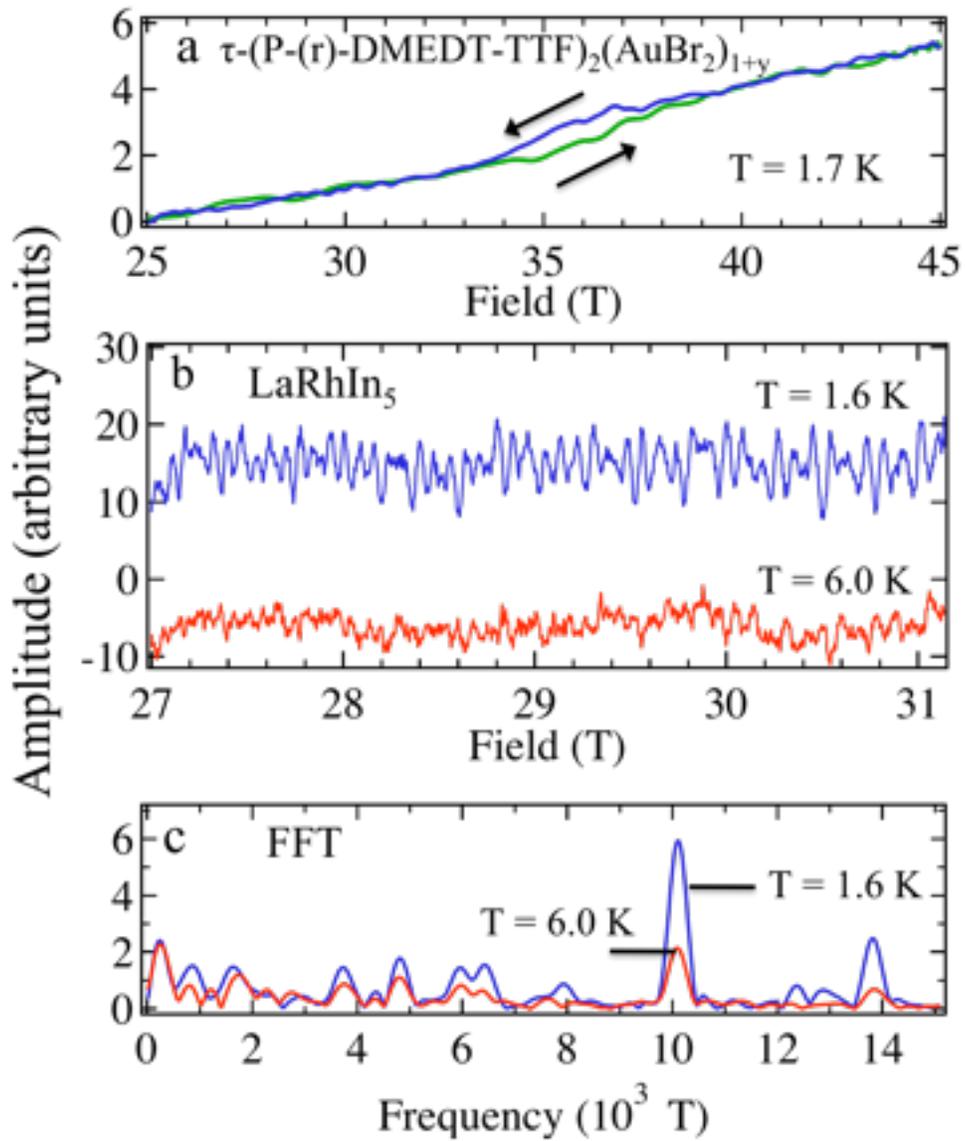

Fig. 4. Sensitivity of PRC dilatometer in pulsed (a) and DC magnetic fields (b). a. magnetostriction of a organic compound, τ-(P-(r)-DMEDT-TTF)$_2$(AuBr$_2$)$_{1+y}$ was measured in a pulsed magnetic field at 1.7 K and shows a metal to insulator transition around 36 T. b. quantum oscillations of a heavy fermion compound, LaRhIn$_5$ measured in DC magnetic fields at 1.6 K and 6 K. c. Fast-Fourier transforms of the quantum oscillations presented in panel b.